\documentclass[12pt]{article}
\usepackage{epsfig}

\def\pr#1{#1^\prime}

\relax
\def\pl#1#2#3{{\it Phys. Lett. }{\bf #1}(19#2)#3}
\def\zp#1#2#3{{\it Z. Phys. }{\bf #1}(19#2)#3}
\def\prl#1#2#3{{\it Phys. Rev. Lett. }{\bf #1}(19#2)#3}
\def\rmp#1#2#3{{\it Rev. Mod. Phys. }{\bf#1}(19#2)#3}
\def\prep#1#2#3{{\it Phys. Rep. }{\bf #1}(19#2)#3}
\def\pr#1#2#3{{\it Phys. Rev. }{\bf #1}(19#2)#3}
\def\np#1#2#3{{\it Nucl. Phys. }{\bf #1}(19#2)#3}
\def\sjnp#1#2#3{{\it Sov. J. Nucl. Phys. }{\bf #1}(19#2)#3}
\def\app#1#2#3{{\it Acta Phys. Polon. }{\bf #1}(19#2)#3}

\relax

\catcode `@ 11
\def\biblabel#1{\if@filesw\immediate
\write\@auxout{\string\bibcite{#1}{\the\value{\@listctr }}}\fi}
\catcode `@ 12
\def    \beq             {\begin{equation}}
\def    \eeq             {\end{equation}}
\def    \beqn             {\begin{eqnarray}}
\def    \eeqn             {\end{eqnarray}}

\def    \=              {\;=\;}
\def    \frac           #1#2{{#1 \over #2}}

\def \as   {\ifmmode \alpha_s \else $\alpha_s$ \fi}

\def \bz   {b_0}

\def \pt   {\mbox{$p_{\scriptscriptstyle T}$}}

\def \mt   {\ifmmode m_{\rm t} \else $m_{\rm t}$ \fi}

\def \to   {\mbox{$\rightarrow$}}

\newcommand\MSB{\ifmmode \overline{\rm MS} \else $\overline{\rm MS}$ \fi}
\newcommand{\ccaption}[2]{
  \begin{center}
    \parbox{0.85\textwidth}{
      \caption[#1]{\small\it {#2}}}
  \end{center}    }

\newcommand\lambdamsb{\ifmmode
\Lambda_5^{\rm \scriptscriptstyle \overline{MS}}
\else
$\Lambda_5^{\rm \scriptscriptstyle \overline{MS}}$
\fi}

\begin{document}
\topskip 2cm
\begin{titlepage}
{\leftskip 11cm
\normalsize
\noindent
\newline
CERN-TH/96-204 \\
hep-ph/9607430

}
\vspace{1cm}
\begin{center}
{\large\bf SOFT GLUON RESUMMATION \\
      IN HEAVY FLAVOUR PRODUCTION}\footnote{Talk
presented at the 10$^{\rm th}$ Rencontres de Physique
  de la Vall\'ee d'Aoste, La Thuile, Val d'Aosta, March 3-9, 1996.} \\

\vspace{1.5cm}
{\large Paolo Nason\footnote{On leave of absence from INFN, Sez. di Milano,
Milan, Italy}} \\
\vspace{.5cm}
{\sl CERN, TH Division, Geneva, Switzerland}\\
\vspace{1.5cm}

\begin{abstract}
I discuss the effect of resummation of soft gluons in hadronic production
of high mass systems, and in particular in heavy flavour production.
I show that in widely used $x$-space resummation formulae, spurious terms that
grow factorially in the order of the perturbative expansion are present.
These terms spoil the convergence of the perturbative expansion.
I also show that it is possible to perform the soft gluon resummation
in such a way that these terms are not present. Implications for top
and high mass dijet production at the Tevatron are discussed.
\end{abstract}

\end{center}
\vfill

\noindent
CERN-TH/96-204 \newline
July 1996    \hfill
\end{titlepage}

\newcommand\hrho{\hat{\rho}}
\section{Introduction}
In this talk I will deal with the problem of the resummation of logarithmically
enhanced effects  in the vicinity of
the threshold region in hard hadroproduction processes.
Drell--Yan lepton pair production has been in the past the best
studied example of this sort \cite{oldresummrefs,AEM,VanNeervenDY}.
The threshold region
is reached when the invariant mass of the lepton pair
approaches the total available energy.
A large amount of theoretical
and phenomenological work has been done on this subject.
References \cite{Sterman} and \cite{CataniTrentadue} summarize all the
theoretical progress performed in this field.
Resummation formulae have also been used in estimating heavy flavour
production \cite{Laenen}, \cite{BergerContopanagos}. In this case only a
leading logarithmic resummation formula is known.
Calculations of the next-to-leading logarithms are in progress
\cite{StermanLathuile}.

Here I will not deal with sophisticated higher order
effects calculations. I will instead describe some recent progress
\cite{cmntnp,cmntpl}
in the understanding of how to implement the resummation.

\section{What are soft gluon effects}
\label{sec:softgluons}
\begin{figure}[htbp]
  \begin{center}
    \leavevmode
    \epsfig{file=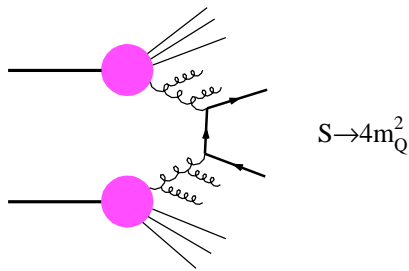,width=6cm}
    \caption{Heavy flavour production near threshold}
    \label{fig:hvqprod}
  \end{center}
\end{figure}

Coloured particles emit soft gluons with high probability.
Normally the effect of soft gluon emission is small (at least in
inclusive quantities), since they only slightly affect the kinematic
of a process. However, in a production process of high mass
objects, when we approach the threshold, soft gluon emission
becomes important. Let us fix our attention on heavy flavour
production in hadronic collisions near threshold.
The process is schematically depicted
in fig.~\ref{fig:hvqprod}.
The incoming protons make a big effort in providing partons
with a large fraction of the longitudinal momentum, thus going towards
the very large $x$ region of the structure functions.
Under these circumstances, even the small amount of energy wasted by soft
gluon radiation yields an important suppression of the cross section.
In the usual application of the factorization theorem, part of the radiative
corrections due to gluon radiation are included in the structure functions.
Thus, depending
upon the factorization scheme, the left-over suppression may be positive
or negative. In the \MSB\ and DIS scheme the left-over is negative,
so that the suppression effect appears instead as an enhancement
of the cross section.

Let us now focus upon the case of heavy flavour production.
The perturbative expansion
for the partonic cross section at order ${\cal O}(\as^3)\,$,
neglecting obvious indices (incoming parton types, etc.),
has the structure
\beq
\hat\sigma(\hat{s})=\sigma_0(\hat{s})\,
\left[1+C\as\log^2(1-\hrho)+{\cal O}(\as\log(1-\hrho),\as^2)\right]\,,
\eeq
where $\hat{s}$ is the square of the partonic centre-of-mass energy,
$\hrho=4m^2/\hat{s}\,$, and $\sigma_0(\hat{s})$ is the Born cross section
\beq
\sigma_0(\hat{s})=\frac{\as^2}{m^2}f^{(0)}(\hrho)\;.
\eeq
The scale at which $\as$ is evaluated is $\as=\as(m^2)$,
unless the argument is explicitly given. 
The reader can find in ref.~\cite{NDE} explicit formulae for the functions
$f^{(0)}(\hrho)$, as well as for the constant coefficient $C$.
The precise value of $C$ depends upon the type of
incoming partons (i.e. quarks or gluons) and upon the factorization
scheme. In both the \MSB\ and the DIS scheme it is positive,
so that in the following we can think of it as being a positive constant.
Resummation, according to ref.~\cite{Laenen}, gives
\beq\label{ResLaenen}
\hat\sigma^{\rm (res)}(\hat{s}) = \frac{\as^2}{m^2}f^{(0)}(\hrho)
\exp\left[C\as(s^\prime)\log^2(1-\hrho)\right]
\eeq
where $s^\prime$ is a scheme-dependent function of $\hat{s}$
that goes to zero as $\hat{s}\,\to\, 4m^2$.
We have $s^\prime=(1-\hrho)^\eta m^2$, where $\eta=1$ in
the DIS scheme, and 3/2 in the \MSB\ scheme.
Formula (\ref{ResLaenen}) is supposed
to include all terms of order $\as^m\log^n (1-\hrho)$ with $n>m$.
Remember in fact that
\beq
\as(s^\prime)=\frac{1}{\bz \log\frac{s^\prime}{\Lambda^2}}
=\frac{1}{\bz \log\frac{m^2}{\Lambda^2}+\bz\eta \log(1-\hrho)}
=\as\left(1-\as\bz\eta\log(1-\hrho)+\ldots\right)\;.
\eeq
so that in the exponent there are terms with arbitrary powers
of $\as$, and a power of the logarithm which is always larger than the power
of $\as$. The first subleading terms have the form
$\as^k\log^k (1-\hrho)$.

In order to get a physical cross section the partonic cross section
given above
should be convoluted with parton luminosities
\beq
\sigma^{\rm (res)}=
\int {\cal L}(\tau)\,\hat\sigma^{\rm (res)}(\tau S)\,d\tau
=\frac{\as^2}{m^2}\int {\cal L}(\tau)f^{(0)}(\hrho)
\exp\left[C\as\left((1-\hrho)^\eta m^2\right)\log^2(1-\hrho)\right]
d\tau
\eeq
where $\hrho=4m^2/(\tau S)$, and (omitting obvious parton indices)
\beq
{\cal L}(\tau)=\int F(x_1)F(x_2)\delta(x_1 x_2-\tau)\,dx_1 dx_2\,.
\eeq
Here we can spot a problem in the resummation formula.
When performing our integral over $\tau$, as $\tau\,\to\,\rho$
the argument of $\as$ in the exponential approaches zero.
Before it actually hits the zero, it will hit
a singularity in the running coupling $\as(s^\prime)$,
causing the integral to diverge. In order to avoid the divergence a cutoff
$\mu_0$ was introduced in the literature \cite{Laenen,Appel}.
Observe that this cutoff has nothing
to do with the standard factorization and renormalization scale $\mu$.
It is essentially a cutoff on soft gluon radiation, imposed in order to avoid
the blowing up of the running coupling associated with soft gluon emission.

The use of a cutoff seems an ad hoc procedure in this case.
It can however be justified to some extent. Suppose, for example,
that we end up in a QCD calculation with a formula like
\beq\label{example}
G=\int_0^{Q^2}\as(k^2) G(k^2) dk
\eeq
where $G(k^2)$ is a smooth function as $k^2\,\to\, 0$.
Integrals of this kind are often found, for example,
in the computation of shape variables in jet physics.
This expression also is divergent as $k^2\,\to\,0$, since at some
point $\as$ approaches the Landau pole. The divergence can be handled
by a cutoff $\mu_0$, which has to be large enough for $\as$ to be barely
perturbative. For example, we may choose $\mu_0=5\Lambda$, a value around
2 GeV. We can then argue that
\beq
G=\int_{\mu_0}^{Q^2}\as(k^2)\, G(k^2)\, dk +C\frac{\mu_0}{Q}\,.
\eeq
In fact, the divergence of the 1-loop expression of $\as$
does not signal a real physical divergence. More likely,
the point at which $\as$ becomes of order 1 signals the breakdown of
perturbation theory. We therefore exclude this region, estimate
its contribution by dimensional analysis, and obtain a power correction.
A slightly more formal justification makes use of the concept of
IR (infrared) renormalons. 
We expand eq.~(\ref{example}) in powers of $\as(Q^2)$, using
\begin{equation}
  \label{asexp}
\as(k^2)=\frac{1}{\bz \log\frac{k^2}{\Lambda^2}}
=\frac{1}{\bz \log\frac{Q^2}{\Lambda^2}+\bz \log\frac{k^2}{Q^2}}
=\as\sum_{j=0}^\infty \left(-\as\bz\log\frac{k^2}{Q^2}\right)^j\;,
\end{equation}
and we get
\begin{eqnarray}
  G&=&\int_0^{Q^2} \as\sum_{j=0}^\infty
  \left( -\as\bz\log\frac{k^2}{Q^2} \right)^j
\nonumber \\ \label{renorms}
 &=& {\as} \sum_{j=0}^\infty (\as\bz)^j\int_0^\infty t^j e^{-t/2} dt
\end{eqnarray}
where $t=\log Q^2/k^2$. The integral can be performed, and one gets
\begin{equation}\label{divexp}
  G=2\as\sum_{j=0}^\infty j!\,(2\as b_0)^j\;,
\end{equation}
which is a divergent series, since a factorial grows faster than any power.
This lack of convergence is in fact a general feature of the perturbative
expansion in field theory. The perturbative expansion should be interpred
as an asymptotic one.
The terms of the expansion (\ref{divexp}) decrease for moderate
values of $j$. As $j$ grows the factorial takes over, the terms
stop decreasing and begin to increase. This happens at the value of $j$
at which the next term is equal to the current one
\beq
(j+1)!\,(2\as b_0)^{j+1}=j!\,(2\as b_0)^j\;
\eeq
or roughly
\beq
j_{\rm min}=\frac{1}{2\as \bz}\;.
\eeq
Asymptotic expansions are usually handled by summing their terms as
long as they decrease. Of course, in this resummation prescription
there is an ambiguity, which is of the order of the size of the first
neglected term. In our case
\beq
j_{\rm min}!(2\as b_0)^{j_{\rm min}}\approx e^{j\log j - j}\frac{1}{j^j}
=e^{-\frac{1}{2\as\bz}}\approx\frac{\Lambda}{Q}\;,
\eeq
which gives a power-suppressed ambiguity with the same power law
that we found using the cutoff procedure.

From the discussion given above, we would expect that the resummation
formulae should include a cutoff of the order of a typical hadronic
scale, and varying the cutoff within a factor of order 1
should affect the cross section by terms of the order $\Lambda/Q$.
In fact, this is not the case.
The cutoff has a dramatic effect on the cross section,
as can be seen from figs.~2 and 3
of ref.~\cite{Laenen}. For example, the uncertainty band obtained
by varying the scale $\mu_0$ between $0.2m$ and $0.3m$ for top
production in the $gg$ channel brings about a change in the cross section
of a factor of 2, for a top mass between 100 and 200~GeV.
These two values of $\mu_0$ correspond to cutting off the gluon
radiation at energies of the order of 20 to 30 GeV, therefore much
larger than a typical hadronic scale.
Other proposals for the resummation procedure have appeared in the
literature. In ref.~\cite{StermanContopanagos} a method was
developed in the context of Drell--Yan pair production,
and it was applied to the heavy flavour case in
ref.~\cite{BergerContopanagos}. Also in ref.~\cite{BergerContopanagos}
(as can be seen from formula (116) and the subsequent discussion)
unphysically large cutoffs are present, much larger than the typical
hadronic scale that one expects.

In the following section, I will show that the presence of
large cutoffs and of large ambiguities in the resummation formula,
is not at all related to the blowing up of the coupling constant. 
In other words, there are other sources of factorial growth
of the perturbative expansion for the resummation of soft gluons,
and they largely dominate the factorial growth due to the running
coupling. I will also show that these terms are spurious, and that
soft gluon resummation can be easily formulated in such a way that these
terms are not present.

\section{Problems with the $x$ space resummation formula}\label{ProbSection}
For definiteness, let us focus upon the resummation
formula (\ref{ResLaenen}). We pointed out earlier that
this formula is divergent when the argument of $\as$ becomes
too small, and the coupling constant blows up. In fact,
formula (\ref{ResLaenen}) is divergent even for fixed coupling
constant. At fixed coupling it can be written as
\beqn
\sigma^{\rm (res)}&=& \frac{\as^2}{m^2}
\int d\tau\; {\cal L}(\tau)f^{(0)}(\hrho)
\exp\left[\as C\log^2(1-\hrho)\right]
\nonumber \\ \label{FixedCoupling}
&=& \frac{\as^2}{m^2} \int d\hrho_\rho^1\;
\frac{\rho}{\hrho^2}{\cal L}\left(\rho/\hrho\right)f^{(0)}(\hrho)
\exp\left[\as C\log^2(1-\hrho)\right]\,,
\eeqn
where $\rho=4m^2/S$,
and the integral diverges as $\hrho\to 1$, since the exponential
\beq
\exp(a\log^2(1-\hrho))=(1-\hrho)^{-a\log(1-\hrho)}
\eeq
grows faster than any inverse power of $1-\hrho$ as $\hrho\,\to\,1$.
This divergence can again be related to factorial growth in the
perturbative expansion.
Expanding formula (\ref{FixedCoupling}) in powers of $\as$
we get 
\beq
\sigma^{\rm (res)}=\frac{\as^2}{m^2}\sum_{k=0}^\infty \frac{1}{k!}(C\as)^k
\int_\rho^1 d\hrho\;
\frac{\rho}{\hrho^2}{\cal L}\left(\rho/\hrho\right)
f^{(0)}(\hrho)\log^{2k}(1-\hrho)\;.
\eeq
It is easy now to see that, due to its singularity for $\hrho\to 1$
the integral of $\log^{2k}(1-\hrho)$ grows like $(2k)!\,$.
Let us make here the simplifying assumption that
$f^{(0)}(\hrho)\approx\theta(1-\hrho)\,$. In a neighbour of the singularity,
the integral has the form
\beq
\int_{1-\epsilon}^1 d\hrho \, \log^{2k}(1-\hrho)
=\int_{\log 1/\epsilon}^\infty  dt\, e^{-t} t^{2k}\;.
\eeq
where we have performed the substitution $t=\log 1/(1-\hrho)\,$.
The above integral equals
\beq
\int_{\log1/\epsilon}^\infty dt\; e^{-t} t^{2k}
=\int_0^\infty dt \; e^{-t} t^{2k}-\int_0^{\log1/\epsilon}  dt
e^{-t} t^{2k} = (2k)!\;
-\int_0^{\log 1/\epsilon}  dt\,
e^{-t} t^{2k}\; .
\eeq
Since
\beq
\int_0^{\log 1/\epsilon}  dt\,
e^{-t} t^{2k} < (\log 1/\epsilon)^{2k+1}
\eeq
we see that the contribution to the integral near the singularity
is dominated by the term $(2k)!\,$. The power
expansion for the cross section is then
\begin{equation}\label{factgr}
\sigma^{\rm (res)}\approx \sum_{k=0}^\infty \frac{(2k)!}{k!}
\left( C\as\right)^k \approx
\sum_{k=0}^\infty k!
\left(4  C\as\right)^k \;.
\end{equation}
The above formula is in fact appropriate for the case of heavy flavour cross
section with a lower cut on the invariant mass of the pair\footnote{A more
accurate analysis, using the known behaviour of $f^{(0)}$ near threshold
would have yielded an extra factor of 4/9 in front of $C$ in
eq.~(\ref{factgr}).}.
As in the previous example, the factorial growth of formula~(\ref{factgr})
will give rise to ambiguitites in the resummation of the perturbative
expansion. These ambiguities are not, however, related to renormalons,
since they occur also at fixed coupling constant. In fact, they are in
general fractional powers of $\Lambda/Q$, where $Q$ is the scale
involved in the problem, and $\Lambda$ is a typical hadronic scale.
For example, in the case of heavy flavour production
at fixed invariant mass of the heavy quark pair via the gluon-fusion
mechanism, the ambiguity would have the form $(\Lambda/Q)^{0.16}$,
and it would thus be extremely relevant even for very massive
heavy flavour pairs.

These large terms in the perturbative expansion are in fact spurious.
They are an artefact of the $x$ space resummation procedure. 
This can be easily understood with the following argument.
Exponentiation of the gluon emission is possible because,
roughly speaking, each soft gluon is emitted independently.
This independence is however only approximate: the total momentum
must be conserved. Momentum conservation, however, is a subleading
effect in the soft resummation formula. Yet, its violation leads
to factorially growing terms. These terms are subleading from
the point of view of the logarithmic behaviour, but very important
from the point of view of the factorial growth of the perturbative expansion.
The presence of large factorial terms due to momentum non-conservation
can be understood also by simple arguments. The emission of $k$
gluons, where each gluon has a limit on its energy $E_i<\eta$,
leads to a phase space that is larger by a factor of $k!$
than the case when the total energy of emission is bounded
$\sum E_i<\eta$. Thus phase space alone provides a $k!$ factor.
We can see in more detail the origin of the $(2k)!\,$ term
by considering the formula for the partonic cross section
with two emitted soft gluons, implementing momentum conservation.
We have
\beq\label{twosoft}
\sigma^{(2)}(\hrho)=\frac{1}{2}(2C\as)^2\frac{\as^2}{m^2}
\int \left[\frac{\log(1-z_1)}{1-z_1}\right]_+
     \left[\frac{\log(1-z_2)}{1-z_2}\right]_+
f^{(0)}({\hrho}^\prime)\,\delta(\hrho-z_1 z_2{\hrho}^\prime)\,
dz_1 dz_2 d{\hrho}^\prime\,.
\eeq
The leading logarithmic term of the above integral is given by
\beq
\sigma^{(2)}(\hrho)=\frac{1}{2}(2C\as)^2\frac{\as^2}{m^2}
\log^4(1-\hrho)\,f^{(0)}(\hrho)+\ldots\,,
\eeq
where terms with less than 4 powers of logarithms are neglected.
We now see that the integral of the leading logarithmic term
of $\sigma^{(2)}(\hrho)$ in $\hrho$ has a large factor $\approx 4!$
due to the integral of $\log^4(1-\hrho)$, while the integral
of the full expression, eq.~(\ref{twosoft}), gives
\begin{equation}
  \label{fullint}
  \int_0^1 d\hrho \sigma^{(2)}(\hrho)=
\frac{1}{2}(2C\as)^2\frac{\as^2}{m^2}\left(\int_0^1 dz
\left[\frac{\log(1-z)}{1-z}\right]_+\right)^2 \int_0^1 d{\hrho}^\prime
f^{(0)}(\hrho)\;=0\,,
\end{equation}
due to the vanishing of the integrals of the soft emission
factors with the $+$ prescription.
In general, we see that if we take generic moments of $\sigma^{(k)}(\hrho)$
(i.e. $\int \hrho^m d\hrho \,\sigma^{(k)}(\hrho)$),
the leading log expression grows like $(2k)!\,$, while the full expression
grows only geometrically with $k$.

The criticism described so far applies to the calculations
of soft gluon effects in heavy flavour production given in refs.~\cite{Laenen}
and \cite{BergerContopanagos}. As one may expect, the large factorial
terms give rise to large corrections to the cross section.
Since the term of the perturbative expansion grow strongly with the
order, they also give large uncertainties. In ref.~\cite{Laenen},
the presence of large uncertainties is in fact recognized.
In ref.~\cite{BergerContopanagos} it is claimed that the uncertainties are
small. Even there, however, an unphysically large cutoff is needed
in order to make sense out of the resummation formulae.

A second, more subtle criticism of the resummation formulae has to do with
the presence of $1/Q$ corrections that arise from infrared renormalons.
It was shown in ref.~\cite{BenekeBraun} that soft gluon resummation
does not yield the correct renormalon structure of the Drell--Yan cross
section. This proof was given in a simplified framework in which
the renormalon structure can be computed exactly.
This result suggests the absence of $1/Q$ corrections in Drell--Yan
cross sections, an issue that is still much debated in the literature
\cite{KorchemskiSterman}--\cite{AkhouryZacharov2}.
We would like to remark, however, that even if $1/Q$ corrections were present
in Drell--Yan and heavy flavour production, they would
only be of the order of 1\% for top production at the Tevatron.

It is possible to formulate the resummation
of soft gluons in such a way that the kinematic constraints are
explicitly satisfied, and no factorial growth arises
in the perturbative expansion. It is enough to formulate
the resummation problem in the Mellin transform space\footnote{In fact,
  resummation formulae are usually derived in Mellin space.}.

In ref.~\cite{cmntnp} we specify a resummation
prescription that uses the Mellin space formula. We demonstrate that
with such prescription there are no factorially growing terms
in the resummed perturbative expansion. Yet, this prescription
consistently includes soft effects. We call this prescription
``Minimal Prescription'' (MP from now on), because it does not
introduce large terms that are not justified by the soft gluon
approximation.

\section{Phenomenological applications}
We have computed various heavy flavour production cross sections
using our MP formula. Details are given in 
refs.~\cite{cmntnp,cmntpl}.
Here I report the most salient results.

\begin{figure}[htb]
\centerline{\psfig{figure=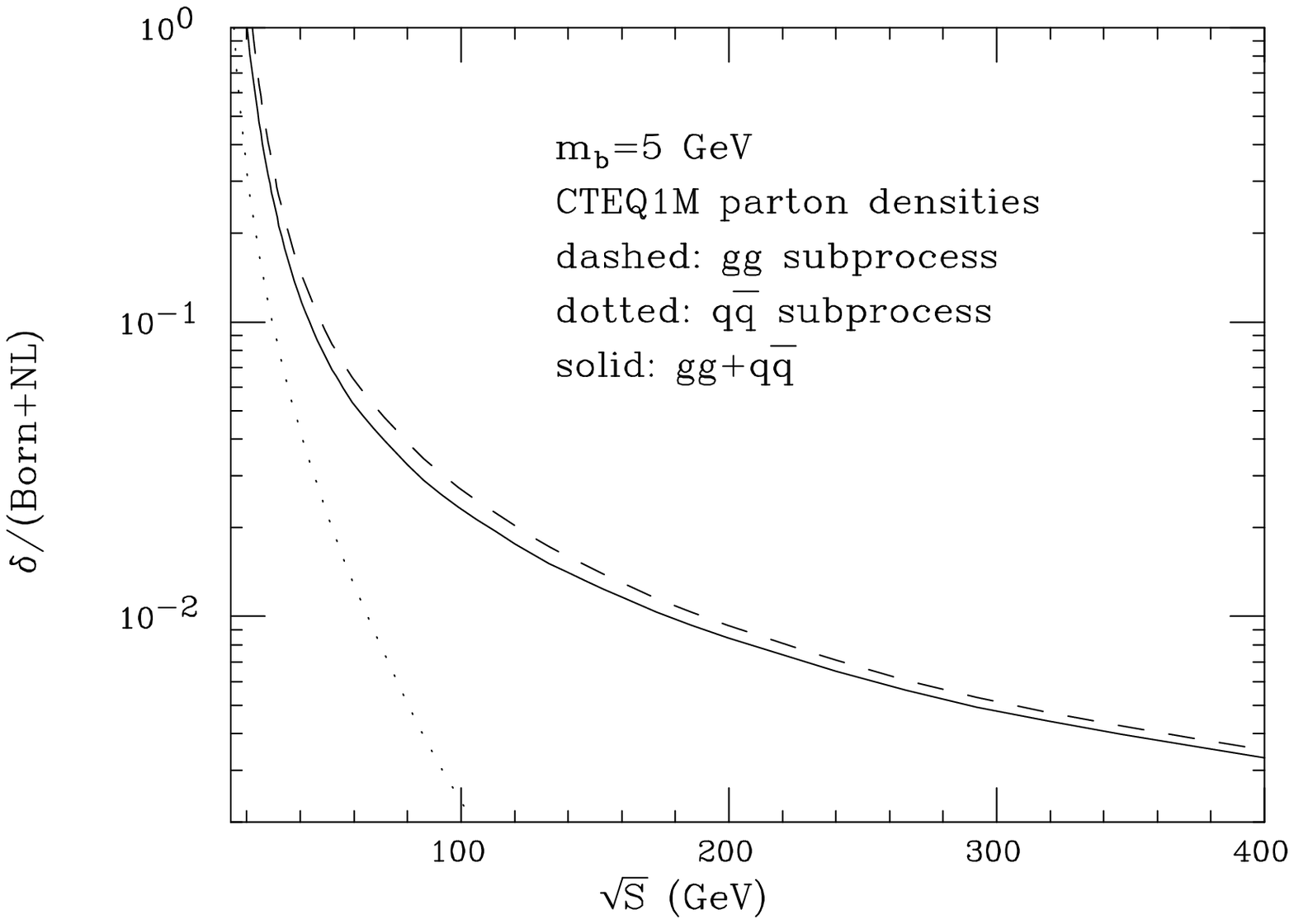,width=0.7\textwidth,clip=}}
\ccaption{}{ \label{bot-kfac}
Contribution of gluon resummation at order $\as^4$ and higher, relative to the
NLO result, for the individual channels and for the total,
for bottom production
as a function of the CM energy in $pp$ collisions.}
\end{figure}

\begin{figure}
\centerline{\epsfig{figure=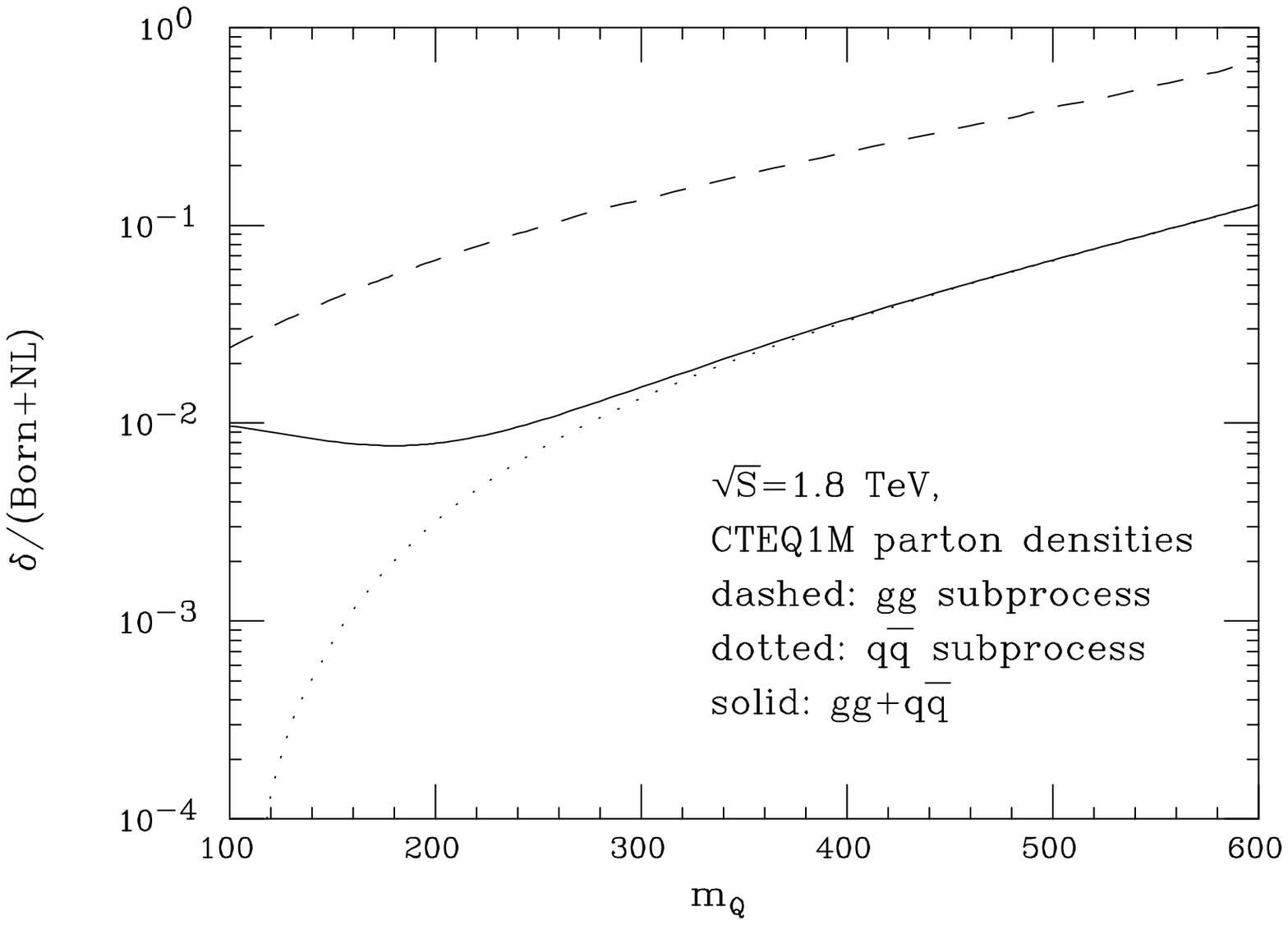,width=0.7\textwidth,clip=}}
\ccaption{}{ \label{frtev}
Contribution of gluon resummation at order $\as^4$ and higher, relative to the
NLO result, for the individual subprocesses and for the total,
as a function of the top mass in $p\bar p$ collisions at 1.8 TeV. }
\end{figure}

\begin{figure}
\centerline{\epsfig{figure=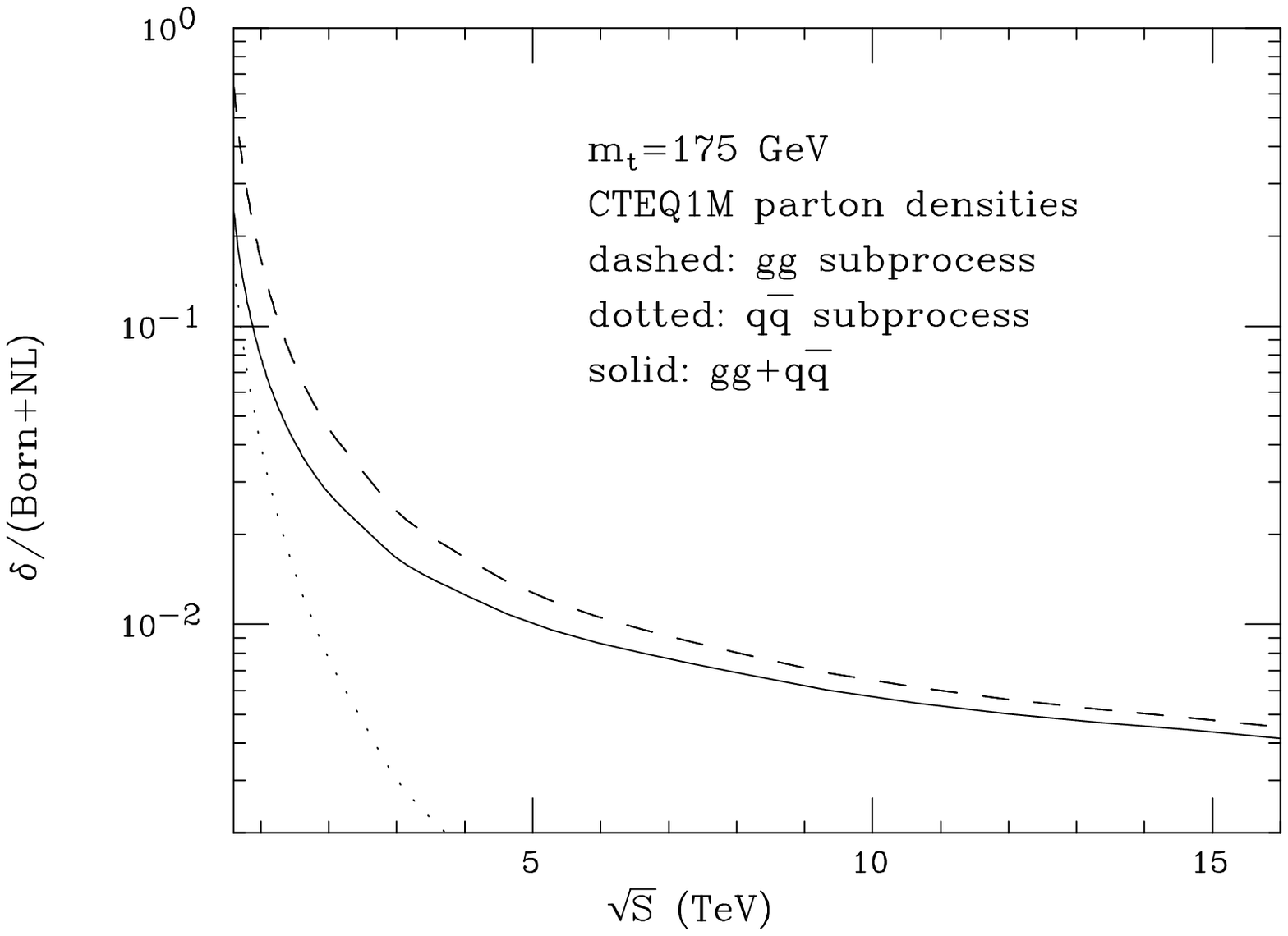,width=0.7\textwidth,clip=}}
\ccaption{}{ \label{frlhc}
Contribution of gluon resummation at order $\as^4$ and higher, relative to the
NLO result, for the individual channels and for the total,
as a function of the CM energy in $pp$ collisions. }
\end{figure}

The importance of the resummation effects is
illustrated in figs.~\ref{bot-kfac}, \ref{frtev} and \ref{frlhc},
where we plot the quantities
\beq \label{deltadef} \frac{\delta_{\rm
    gg}}{\sigma^{(gg)}_{\rm NLO}}\,,\quad \frac{\delta_{\rm
    q\bar{q}}}{\sigma^{(q\bar{q})}_{\rm NLO}}\,,\quad
\frac{\delta_{\rm gg}+\delta_{\rm q\bar{q}}}{\sigma^{(gg)}_{\rm
    NLO}+\sigma^{(q\bar{q})}_{\rm NLO}}\;.
\eeq
Here $\delta$ is equal
to our MP-resummed hadronic cross section in which the terms of order
$\as^2$ and $\as^3$ have been subtracted, and $\sigma_{(\rm NLO)}$ is
the full hadronic NLO cross section.
The results for $b$ at the Tevatron can be easily inferred
from fig.~\ref{bot-kfac}, since the $q\bar{q}$ component is negligible
at Tevatron energies.

For top production, we see that in most configurations of practical
interest, the contribution of resummation is very small, being of the order
of 1\% at the Tevatron.
A complete review of top quark production at the Tevatron,
based upon these findings, has already been given in
ref.~\cite{cmntpl}. We also observe that, for top production at the LHC,
soft gluon resummation effects are negligible. Of course, in this last case,
there are other corrections, not included here, that may need to be
considered. Typically, since the values of $x$
involved are small in this configuration, one may have to worry
about the resummation of small-$x$ logarithmic effects \cite{smallx}.

We see from the figures that in most experimental configurations of
interest these effects are fully negligible. One noticeable
exception is $b$ production at HERAb, at $\sqrt{S}=39.2$, where
we find a 12\% increase in the cross section.
%
%
%
This correction is however well below the uncertainty due to higher order
radiative effects. For example, from the NLO calculation
with the MRSA$^\prime$ \cite{MRSAp} parton
densities and $m_b=4.75\,$GeV, we get
$\sigma_{b\bar{b}}=10.45{ +8.24  \atop -4.65}\,$nb,
a range obtained by varying the renormalization
and factorization scales from $m_b/2$ to $2m_b$. Thus
the upper band is 80\% higher than the central value, to be compared with
a 10\% increase from the
resummation effects. This result is much less dramatic than
the results of ref.~\cite{KidonakisSmith95}.

\section{Jet Cross Sections}\label{jcs}
The interest in the effects of resummation on the behaviour of jet cross
sections at large energy is prompted by the discrepancy between the
single-inclusive jet-\pt\ distribution at large \pt, as measured by CDF
\cite{cdfjet}, and the result of the NLO QCD predictions \cite{nlojet}.  For
simplicity we will study the effects of soft gluon resummation on the
invariant mass distribution of the jet pair, which is, from a theoretical
point of view, very close to the Drell--Yan pair production.  Observe that
other distributions, such as the $\pt$ of the jet, have a rather different
structure from the point of view of soft gluon resummation. In fact, while the
jet pair mass is only affected by the energy degradation due to initial state
radiation, the $\pt$ of the jet may also be affected by the transverse
momentum generated by initial state radiation, and by the broadening of the
jet due to final state radiation.

A study of the jet pair mass distribution is not of purely academic interest,
since also for this variable an analogous discrepancy between data and theory
has been observed \cite{cdfmass}.  Studies of resummation effects in the
inclusive $\pt$ spectrum of jets are in progress (M. Greco and P. Chiappetta,
private communication).

\begin{figure}
\centerline{\epsfig{figure=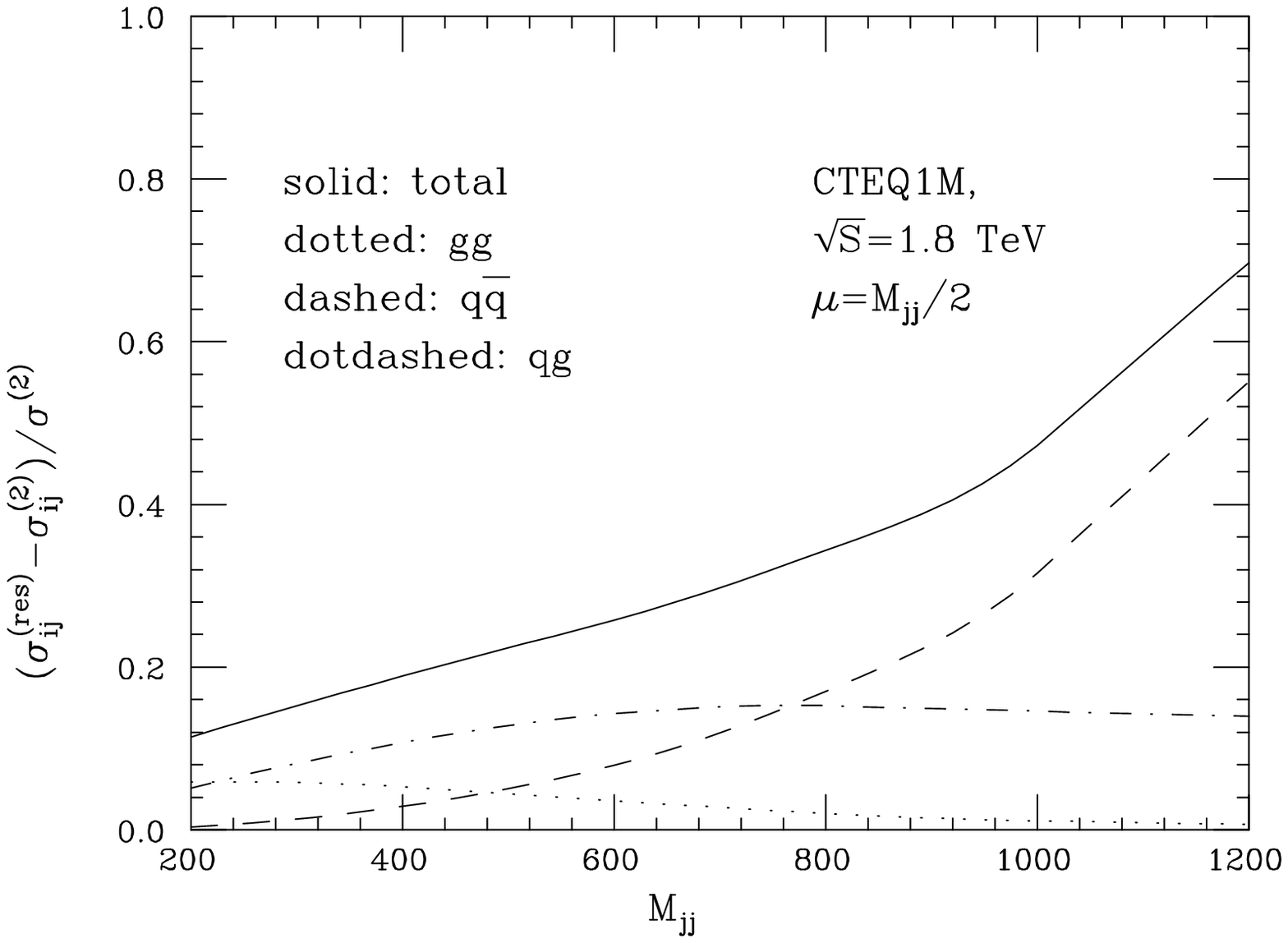,width=0.66\textwidth,clip=}}
\ccaption{}{ \label{jetcteq1}
Contribution of gluon resummation at order $\as^3$ and higher, relative to the
LO result, for the invariant mass distribution of jet pairs at the Tevatron.}
\vskip 1cm
\centerline{\epsfig{figure=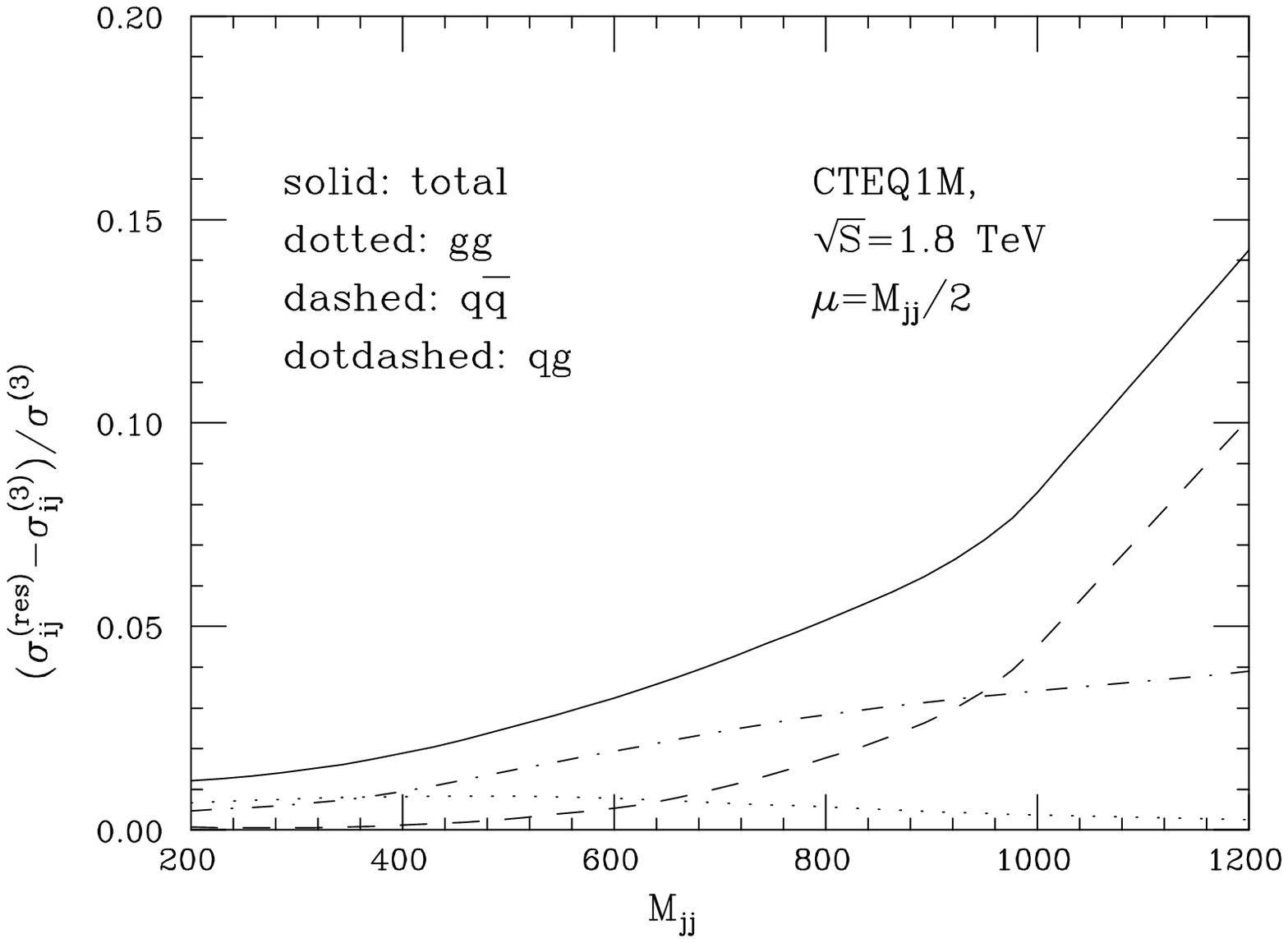,width=0.66\textwidth,clip=}}
\ccaption{}{ \label{jetcteq2}
Contribution of gluon resummation at order $\as^4$ and higher, relative to the
truncated ${\cal O}(\as^3)$ result,
for the invariant mass distribution of jet pairs at the Tevatron.}
\end{figure}

In fig.~\ref{jetcteq1} we show the following quantities:
\beq
\frac{\delta^{(3)}_{\rm gg}}{\sigma^{(2)}}\,,\quad
\frac{\delta^{(3)}_{\rm qg}}{\sigma^{(2)}}\,,\quad
\frac{\delta^{(3)}_{\rm q\bar{q}}}{\sigma^{(2)}}\,,\quad
\frac{\delta^{(3)}_{\rm gg}+\delta^{(3)}_{\rm qg}+\delta^{(3)}_{\rm
q\bar{q}}}{\sigma^{(2)}}
\eeq
where $\delta^{(3)}$ is equal to
our MP resummed hadronic cross section
in which the terms of order $\as^2$
have been subtracted, and
$\sigma^{(2)}=\sigma^{(2)}_{\rm gg}+\sigma^{(2)}_{\rm qg}
 +\sigma^{(2)}_{\rm q\bar{q}}$
is the full hadronic LO cross section
(of order $\as^2$).
We use as a reference renormalization and factorization
scale for our results $\mu=M_{jj}/2$. Notice that for
large invariant masses the effects of higher orders are large.
To understand how much is due to the first order corrections
(which are exactly calculable \cite{nlo2jet})
and how much is due to corrections of order
$\as^4$ and higher, we show the following quantities in fig.~\ref{jetcteq2}
\beq
\frac{\delta^{(4)}_{\rm gg}}{\sigma^{(3)}}\,,\quad
\frac{\delta^{(4)}_{\rm qg}}{\sigma^{(3)}}\,,\quad \frac{\delta^{(4)}_{\rm
    q\bar{q}}}{\sigma^{(3)}}\,,\quad \frac{\delta^{(4)}_{\rm
    gg}+\delta^{(4)}_{\rm qg}+\delta^{(4)}_{\rm q\bar{q}}}{\sigma^{(3)}} \; ,
\eeq
where $\delta^{(4)}$ is now equal to the MP resummed hadronic cross section
with terms of order $\as^3$ subtracted, and $\sigma^{(3)}$ is an approximation
to the full NL cross section, summed over all subprocesses, obtained by
truncating the resummation formula at order $\as^3$.  This figure shows that
indeed most of the large $K$ factor is due to the pure NLO corrections, with
the resummation of higher order soft gluon effects contributing only an
additional 10\% at dijet masses of the order of 1~TeV.

These results
should only be taken as an indication of the order of magnitude of the
correction, since we have not included here a study of the resummation
effects on the determination of the parton densities.  From this
preliminary study it seems however unlikely that the full 30--50\%
excess reported by CDF for jet \pt's in the range 300--450~GeV could
be explained by resummation effects in the hard process. It is
possible that the remaining excess be due to the poor knowledge of the
gluon parton densities at large $x$, an idea pursued by the CTEQ group
\cite{tung}.
\clearpage
\relax
\def\pl#1#2#3{{\it Phys. Lett. }{\bf #1}(19#2)#3}
\def\zp#1#2#3{{\it Z. Phys. }{\bf #1}(19#2)#3}
\def\prl#1#2#3{{\it Phys. Rev. Lett. }{\bf #1}(19#2)#3}
\def\rmp#1#2#3{{\it Rev. Mod. Phys. }{\bf#1}(19#2)#3}
\def\prep#1#2#3{{\it Phys. Rep. }{\bf #1}(19#2)#3}
\def\pr#1#2#3{{\it Phys. Rev. }{\bf #1}(19#2)#3}
\def\np#1#2#3{{\it Nucl. Phys. }{\bf #1}(19#2)#3}
\def\sjnp#1#2#3{{\it Sov. J. Nucl. Phys. }{\bf #1}(19#2)#3}
\def\app#1#2#3{{\it Acta Phys. Polon. }{\bf #1}(19#2)#3}

\end{document}